\documentclass[aip, amsmath, amssymb, reprint]{revtex4-1}

\usepackage[utf8]{inputenc}
\usepackage[english]{babel}

\usepackage{graphicx}
\usepackage{dsfont}
\usepackage{dcolumn}
\usepackage{bm}
\usepackage{color}
\usepackage{xcolor}
\usepackage{url}
\usepackage{tabularx}
\usepackage{array}
\usepackage{booktabs}
\usepackage{comment}
\usepackage[colorlinks=true]{hyperref}
\hypersetup{linkcolor=[rgb]{0.6 0 0}, citecolor=[rgb]{0 0.4 0}, urlcolor=[rgb]{0 0 0.6}}
\usepackage{lipsum}
\usepackage[normalem]{ulem}
\usepackage{titlesec}

\begin{document}

\title{The absence of superconductivity in the next-to-leading order Ginzburg-Landau functional for Bardeen-Cooper-Schrieffer superconductor}

\author{Filipp N. Rybakov}
\email{prybakov@kth.se}
\affiliation{Department of Physics, KTH Royal Institute of Technology, SE-10691 Stockholm, Sweden}

\author{Egor Babaev}
\affiliation{Department of Physics, KTH Royal Institute of Technology, SE-10691 Stockholm, Sweden}

\begin{abstract} 
Shortly after the Gor'kov microscopic derivation of the Ginzburg-Landau (GL) model via a small order parameter expansion in Bardeen-Cooper-Schrieffer theory of superconductivity, the derivation was carried to next-to-leading order in that parameter and its spatial derivatives. The aim was to  obtain a generalized GL free energy that approximates the microscopic model better. Since 1960s, multiple works have claimed or implicitly assumed that this extended GL model corresponds to the free energy and has solutions in the form of local minima describing superconductivity, such as vortex solutions. In contrast to this, we prove that this extended GL functional does not represent free energy since it does not have any solutions in the form of minima. Accordingly, it cannot be used to describe superconducting states. 
\end{abstract}

\maketitle

\section{Introduction}

The Ginzburg-Landau (GL) model of superconductivity has been and continues to be an extremely useful tool. Retaining important degrees of freedom it allows one to describe and analyze inhomogeneous states at a large length scale where no analytical and no numerical solutions of  microscopic models are available.
Gor'kov provided a microscopic derivation of the GL functional from the Bardeen-Cooper-Schrieffer (BCS) model~\cite{Gorkov1959}.
Namely, he has shown that GL model emerges by inducting the complex order parameter field ${\psi(\mathbf{r})}$, which is proportional to superconducting gap function ${\Delta(\mathbf{r})}$, and leading order expansion in small amplitude and small gradient of this order parameter. 
Almost immediately afterward, the expansion was carried to the next-to-leading order~\cite{tewordt1963gap,tewordt1964generalized,tewordt1965generalized,neumann1966structure,neumann1966structure,Eilenberger1966,jacobs1971theory}.  
In what follows, we refer to this as the extended GL model.

The aim of the extended GL model is to more accurately approximate the microscopic theory. 
In this regard, it is worth mentioning the issue of vortex interaction in the regime of the GL parameter near the Bogomolny  point, ${\kappa\approx 1/\sqrt{2}}$.
In the standard GL model the vortices do not interact when ${\kappa=1/\sqrt{2}}$.
However, the solution obtained in the Eilenberger model shows that there are microscopic corrections that lead to a weak non-monotonic interaction that extends up to ${\kappa \lessapprox 1.1}$ at ${T\to 0}$~\cite{Eilenberger,klein, Weber_1987}, and vanishes close to critical temperatures ${T\to T_\text{c}}$,  consistently with  the standard picture that in the GL model vortices interact attractively for ${\kappa < 1/\sqrt{2}}$ or repulsively for ${\kappa > 1/\sqrt{2}}$. 
The Eilenberger equations are a microscopic model that retains more degrees of freedom of BCS theory compared to the GL model and  do not rely on the expansion of the order parameter. 
In addition, the validity of Eilenberger equations is not restricted to the vicinity of critical temperature. 
In this regard, the question was raised if an extension of the GL model may bring the results closer to those obtained in the quasi-classical Eilenberger formalism.

However the problem that arises in the next-to-leading order expansion of BCS theory is the alternation of signs of the coefficients which in turn means that the GL functional deduced from the expansion-derived equations is unbounded from below. 
To verify this, it suffices to consider ${|\psi|\rightarrow\infty}$ in~Ref.\cite{neumann1966structure} or, equivalently, ${|\Delta|\rightarrow\infty}$ in Ref.~\cite{ovchinnikov1999generalized}.
This fact that there are no global minima does not necessarily mean  the absence of solutions in the form of local minima. 
With this in mind, in previous studies the extended GL functional was interpreted as free energy and various energy-based arguments were suggested to estimate length scales and vortex properties~\cite{neumann1966structure,neumann1966structure,jacobs1971theory,jacobs1971first,jacobs1971interaction,HubertAttractive1972,ovchinnikov1999generalized,Ovchinnikov2013,luk2001theory,vagov2012extended, vagov2016superconductivity,wolf2017vortex,vagov2020universal}.
Here we prove that the assumption of existence of metastable states is incorrect.
Therefore, within the GL formalism expanded to the next-to-leading order there is no superconductivity since there are no stable solutions corresponding to the superconducting state, i.e. the system does not have the Meissner effect and vortices.

\section{The model}

Consider the next-to-leading order Ginzburg-Landau family of models for a BCS superconductor~\cite{tewordt1963gap,tewordt1964generalized,tewordt1965generalized,neumann1966structure,jacobs1971first,jacobs1971interaction,jacobs1971theory,ovchinnikov1999generalized,Ovchinnikov2013,vagov2012extended,vagov2016superconductivity}:
\begin{align}
f\ =\ &C_1\,\mathbf{B}^2 + 
C_2\,|\mathbf{D}\,\psi|^2 - C_3 (|\psi|^2 - \rho_0^2) +  \nonumber \\ 
& C_4 (|\psi|^4 - \rho_0^4) - p_1 (|\psi|^6 - \rho_0^6)  -  p_2\,|\psi|^2 |\mathbf{D}\,\psi|^2 - \nonumber \\
& p_3 \left( (\psi{\mathbf{D}^{*}\,\psi^{*}})^2 + (\psi^{*}\mathbf{D}\,\psi)^2 \right) - p_4\,|\mathbf{D}^2\,\psi|^2 - \nonumber \\
& p_5\,\mathrm{curl}\,\mathbf{B}\cdot\mathbf{i} - p_6\,\mathbf{B}^2 |\psi|^2,
\label{f} 
\end{align}
where $f$ is considered as the free energy density, and 
\begin{align}
\mathbf{B} &= \mathrm{curl}\,\mathbf{A}, \\
\mathbf{D} &= \mathbf{\nabla} + i\,q\,\mathbf{A}, \\
\mathbf{i} &= \text{Im}( \psi^{*}\mathbf{D}\,\psi ).
\label{model}
\end{align}
In our notation, the positive coefficients $C_{1,2,3,4}$ denote whatever notations of the standard GL expression, while positive coefficients $p_{1,2,3,4,5,6}$ denote additional  contributions derived from the BCS model as higher order expansion terms.
The microscopically derived coefficients can, for example, be found in Refs.~\cite{neumann1966structure,ovchinnikov1999generalized}.
Without loss of generality, we choose an additive constant in such a way that the potential term gives zero at its minimum, namely 
\begin{align}
\rho_0 = \sqrt{  \frac{ C_4 - \sqrt{C_4^2 - 3 C_3 p_1} }{3 p_1}  }. \label{rho0}
\end{align}
Therefore, for example, in the limit ${p_1 \to 0}$, Eq.~(\ref{rho0}) gives standard GL density ${\rho_0 = \sqrt{C_3/2 C_4}}$.

The constants $p_k$ are material-dependent and temperature-dependent. It is important that, as long as one stays within the standard BCS theory, these
constants do not vanish, including at BCS critical temperature, $T_\text{c}$. 
Let us estimate the minimum value of $p_4$. In the limit of a clean superconductor, from Refs.~\cite{ovchinnikov1999generalized,luk2001theory} it immediately follows that
\begin{align}
\frac{p_4}{C_2} = 
\frac{93 \zeta(5)}{ 560 \zeta(3) \pi^2}
\left(
\frac{\hbar v_\text{F}}{k_\text{B} T_\text{c}} 
\right)^2 
\left(\frac{T_\text{c}}{T}
\right)^2.
\end{align}
By taking into account that the BCS coherence length, ${\xi_\text{BCS}\approx 0.18 \hbar v_\text{F}/k_\text{B} T_\text{c}}$, is finite, we find
\begin{align}
p_4 \gtrapprox 0.45\, C_2 \, \xi_\text{BCS}^2 , \label{p4_min}
\end{align}

In the similar way we obtain that 
\begin{align}
& p_2 \gtrapprox 2\, C_4 \, \xi_\text{BCS}^2 ,  \label{p2_min}\\
& p_3 \gtrapprox 0.25\, C_4 \, \xi_\text{BCS}^2 . \label{p3_min}
\end{align}

\section{Exact proof of the absence of minima, by Legendre condition}

The proof that there are no minima, neither local nor global, irrespective of how close one is to critical temperature, follows from the fact that~(\ref{f}) does not pass the Legendre test~\cite{Forsyth1927} for any field $\psi(\mathbf{r})$, while such a test is a necessary condition for existence of a minimum in a functional. 
Indeed, taking into account that $f$ depends on ${\text{Re}\,\psi}$ and its spatial derivatives up to the second order, we immediately obtain that 
\begin{align}
\frac{\partial^2 f}{\partial \left(  \partial_{xx}\text{Re}\,\psi \right)^2} = -2 p_4 < 0.
\label{Legendre}
\end{align}
That is, the found second derivative is negative everywhere, which means that there are no solutions for the fields that would provide  a minimum for the functional. Note that for the density of the kind~(\ref{f}) the Legendre test differs (see, for example, Sec.~348 in Ref.~\cite{Forsyth1927}) from the classic Legendre condition known also as Legendre-Hadamard~\cite{GiaquintaHildebrandt} or Clebsch-Legendre condition.
While our demonstration is valid for both finite and infinite systems, for mathematical precision, we state that the space where the functions $\psi$ are defined is assumed to be an open subset of $\mathbb{R}^d$, ${d=1,2~\text{or}~3}$. 
For the case of bounded subsets, we assume the natural boundary conditions for the superconductor-vacuum interface $\Gamma$, i.e. the vanishing of the perpendicular current. 
In turn, by virtue of Maxwell's equations, this means that  
\begin{align}
(\mathrm{curl}\,\mathrm{curl}\,\mathbf{A})_\perp &= 0\quad\text{on}\ \Gamma,
\end{align}
whereas $\psi$ is unconstrained in the formulation of the minimization problem. 
However, we can also generalize our conclusion to the case of a constraint in the form of given $\psi$ on $\Gamma$. 
Anyway, for any $\psi$ on $\Gamma$ the Eq.~(\ref{Legendre}) proves that there is no minimum.

The reason for the absence of a minimum is that for any given field configuration there are always various perturbations, arbitrarily small in amplitude and gradients, which lower the ``energy''.
It is important to note that it does not matter at all whether the Euler-Lagrange equations are satisfied.

\section{Destabilizing infinitesimally small perturbations}

The results of the Legendre test guarantee the absence of a minimum, and hence the existence of infinitely weak perturbations that destabilize any given initial state. 
The scenario of instability consists in a perturbation, for which the energy contribution from the second derivative dominates over the rest of the contributions. 
Here we give explicit examples of a couple of infinitesimal perturbations that lead to a runaway instability even of the ${\psi(\mathbf{r}) = \text{const}}$ configuration. 
Note that these perturbations are not unique. 

\subsection{An example of a  perturbation in the form of a periodic function}

\begin{figure}
    \centering
    \includegraphics[width=7.2cm]{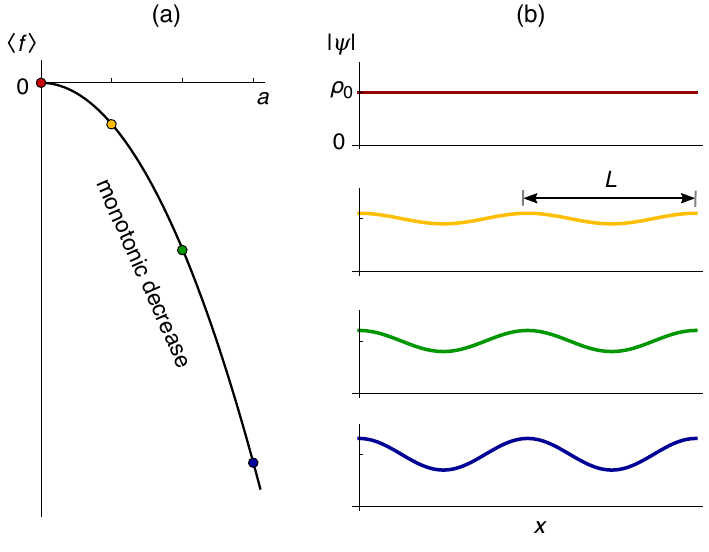}
    \caption{
Infinitesimally weak perturbation with a periodic profile driving the system into a runaway catastrophe. 
(a) Average ``energy'' as a function of shape parameter $a$. 
(b) Spatial field distributions for points marked in (a), in accordance with Eq.~(\ref{psi}).
} \label{fig1}
\end{figure}

Consider a family of fields depending on coordinate $x$ and an additional scalar parameter ${a \geq 0}$ in such a way that the limit $a = 0$ corresponds to a uniform superconducting state (if such a state exists), 
\begin{align}
& \mathbf{A} = 0, \nonumber \\
& \psi = \rho_0 \left(1 + a \cos\left(2\pi x / L \right) \right)\exp( i \phi), \label{psi}
\end{align}
where $\phi$ denotes the arbitrary constant of the phase shift.
The modulus is chosen so that it would correspond to a local minimum of the Landau model without gradient terms.

Substituting~(\ref{psi}) into~(\ref{f}), we find the following average density:
\begin{align}
\langle f \rangle &= \frac{1}{L} \int\limits_{0}^{L} f dx =  
-a^2 \Bigg[  \frac{2\pi^2 C_2 \rho_0^2}{L^4}\left(  \frac{4\pi^2 p_4}{C_2} - L^2 \right) \nonumber \\
& + \frac{2 C_4 \rho_0^4}{L^2} \left( \frac{\pi^2(p_2+2p_3)}{C_4} - L^2 \right) + 6 p_1\rho_0^6 \Bigg]  \nonumber \\
& -a^4 \left[   \frac{3 C_4 \rho_0^4}{8L^2} \left( \frac{4\pi^2(p_2+2p_3)}{3C_4} - L^2 \right) + \frac{45 p_1 \rho_0^6}{8} \right]  \nonumber \\
& -a^6\frac{5 p_1 \rho_0^6}{16}. 
\label{unbound}
\end{align}
Choosing a period in the range
\begin{align}
0 < L \leq 2\pi \sqrt{ \min\left( 
\frac{p_2 + 2 p_3}{4 C_4}, 
\frac{p_4}{C_2}  \right)}, \label{l1}
\end{align}
ensures that the average density of $f$ becomes a monotonically decreasing unbounded function of the parameter ${a \in [0, \infty)}$.
That is, an arbitrarily small perturbation destabilizes the assumed state of superconductivity and results in a blow-up, see Fig.~\ref{fig1}. 

The perturbation scale, $L$, does not play a significant role since the effective theories and free energy functionals require stability regarding infinitesimal perturbations of any scale.  
Note that the range~(\ref{l1}) never vanishes: see~(\ref{p4_min})-(\ref{p3_min}).

\subsection{An example of a localized destabilizing perturbation}
\label{pert_loc}

\begin{figure}
    \centering
    \includegraphics[width=7.2cm]{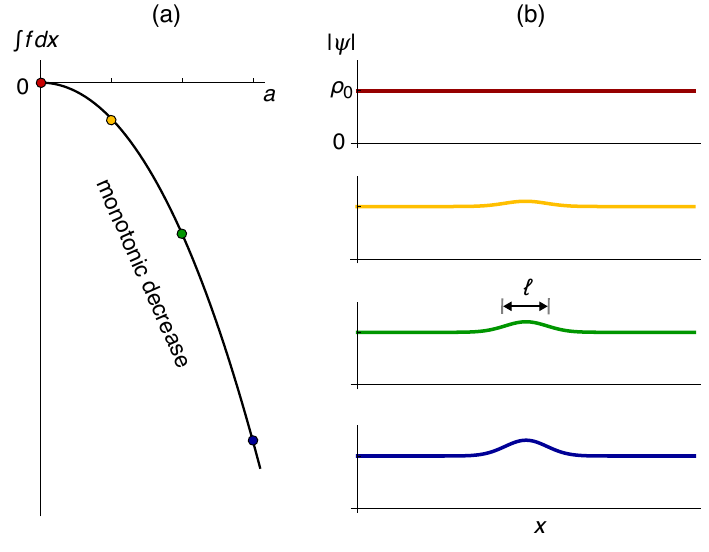}
    \caption{
Infinitesimally weak perturbation, driving the system into a runaway catastrophe. 
(a) Total ``energy'' as a function of shape parameter $a$. 
(b) Spatial field distributions for points marked in (a), in accordance with Eq.~(\ref{psi_loc}).
} \label{fig2}
\end{figure}

Consider the family of the following fields: 
\begin{align}
& \mathbf{A} = 0, \nonumber \\
& \psi = \rho_0 \left(1 + a \exp\left(-2 x^2 / l^2 \right) \right)\exp( i \phi). \label{psi_loc}
\end{align}
Substituting~(\ref{psi_loc}) into ${E = \int f dx}$, and applying an analysis similar to that used for~(\ref{unbound}), we obtain the range for $l$ entailing a blow-up (see Fig.~\ref{fig2}),
\begin{align}
0 < l \leq \sqrt{\min \left( 
\frac{p_2 + 2 p_3}{2 C_4}, 
\frac{6 p_4}{C_2}  \right)}. \label{l2}
\end{align}

\section{Discussion}

The perturbations described in the previous section do not exhaust all their diversity. 
Hence, for example, if one translates~(\ref{f}) into a numerical minimization algorithm, then such an algorithm must lead to blow-up, while the scenarios can be very different between specific implementations and depend on the initial guesses for the fields.

A theory that admits stable or metastable states, should be robust against an infinitely weak point-like perturbation. The length scale at which a system recovers a uniform state defines the coherence length.  
The example from Sec.~\ref{pert_loc} shows that this principle is violated here, and the system has no coherence length.
In this regard, it is useful to discuss separately the approach to~(\ref{f}) using the  expansions near $T_\text{c}$, i.e. by assuming that the parameter $\tau = 1 - T/T_\text{c}$ is small.
Therefore, for example, in Ref.~\cite{neumann1966structure} the applicability of the standard GL scaling is postulated for~(\ref{f}), and the variables are replaced by rescaled ones as follows: $\psi \rightarrow \sqrt{\tau}\psi$, $\mathbf{r} \rightarrow \mathbf{r}/\sqrt{\tau}$, and $\mathbf{A} \rightarrow  \sqrt{\tau}\mathbf{A}$. 
After dropping the common factor $\tau^2$, it effectively changes expression~(\ref{f}) only in that all the coefficients $p_i$ are converted according to $p_i \rightarrow \tau\cdot\text{const}_i$. It is then assumed that due to the smallness of $\tau$, the terms that, in our notation, are multiplied by $p_i$ play the role of corrections. However, this is erroneous due to the fact that the prefactor tending to zero does not necessarily nullify here the total contribution of the corresponding term. 
Indeed, if $\psi$ aims to provide a minimum for the model in question, then 
\begin{align}
\lim_{\tau \rightarrow 0} \left( \tau \int  |\mathbf{D}^2\,\psi|^2 d\mathbf{r} \right) \neq 0,
\end{align}
because in accordance with the Legendre criterion, there is always a degree of freedom for $\psi$ to reshape and decrease the ``energy'' monotonically, therefore providing the value of ${\int  |\mathbf{D}^2\,\psi|^2 d\mathbf{r}}$ arbitrarily large, and, in particular, larger than~$\tau^{-1}$. 
In turn, that implies that here one cannot classify the gradient terms by powers of $\tau$. To summarize, we conclude that the limit ${p_{1,2,...6} \rightarrow 0}$, is a singular limit, and it does not gradually converge to the case when all corresponding terms are absent.
Thus, in the model under consideration, there is no passage to the limit to the standard GL. 
Consequently, the conclusions obtained when considering~(\ref{f}) as the density of free energy (see, for example, Refs.~\cite{vagov2012extended, vagov2016superconductivity,wolf2017vortex,vagov2020universal}) are false irrespective of the proximity to the critical temperature.  
That statement also applies to spurious multiband versions~\cite{vagov2016superconductivity}, which have similar gradient terms but are derived from an erroneous expansion of multiband BCS model, yielding claims of phase diagrams that are principal disagreements with the phase diagrams obtained in microscopic multiband Eilenberger theory~\cite{Silaev.Babaev:11,Silaev.Babaev:12} (see the discussion of these errors in Sec.~4.10 in Ref.~\cite{wordenweber2017superconductorsBabaev}).

\section{Conclusion}

We considered the extended Ginzburg-Landau functional derived by next-to-leading order expansion in the order parameter and its gradients.
We demonstrated that the truncation in the form~(\ref{f}) of the corresponding expansion in the BCS theory leads to ill-posedness and catastrophe, and the expression~(\ref{f}) does not represent free energy density. 
This in contrast to multiple previous studies assumed that the above expression represents free energy density~\cite{neumann1966temperature, vagov2012extended, cavalcanti2020multiband}, and claimed the existence of minima~\cite{ovchinnikov1999generalized, Ovchinnikov2013}.

We note that our conclusions do not apply for spin imbalanced superconductors, where
 higher-order generalizations of the Ginzburg-Landau functional have energy minimizing solutions both for pair-density-wave states~\cite{buzdin1997generalized,Radzihovsky2011,barkman2019antichiral,samoilenka2020microscopic,samoilenka2020pair} and for homogeneous states~\cite{barkman2020ring}.

Our result leads to a natural question if keeping more terms (i.e., next-next-to-leading order and maybe even higher order) in the GL functional would lead to a well-posed problem and, potentially to a better approximation of the BCS theory than the standard GL theory, or such a correction lies beyond all orders. 
The answer to this question is beyond the scope of this work.

\begin{acknowledgments}
We thank Martin Speight and Mihail Silaev for discussions. 
This work was supported by the Swedish Research Council Grants No. 642-2013-7837, 2016-06122, 2018-03659, and G\"{o}ran Gustafsson Foundation for Research in Natural Sciences. 
\end{acknowledgments}

\bibliography{bibliography.bib}

\end{document}